%% file: mplocwcnc.tex
\documentclass[conference,10pt]{IEEEtran}
\IEEEoverridecommandlockouts

\usepackage[T1]{fontenc}
\usepackage[english]{babel}
\usepackage[utf8]{inputenc}
\usepackage{csquotes}

\usepackage{graphicx}
  \usepackage[outdir=./img/]{epstopdf}
   \graphicspath{{./img/}}
   \DeclareGraphicsExtensions{.eps,.pdf,.png,.jpg}   
\usepackage{subfigure}
\usepackage{caption}
\usepackage[inkscapelatex=false]{svg}
\usepackage{ifthen}

\usepackage{tikz}
\usepackage{tikzscale}
\usepackage{tikz-dimline}
\usetikzlibrary{plotmarks,patterns,decorations.pathreplacing,backgrounds,calc,arrows,arrows.meta,spy,matrix,backgrounds,shapes,math}

\tikzset{
    block/.style = {draw, rectangle, 
        minimum height=1cm, 
        minimum width=1.2cm, align=center},
    input/.style = {coordinate,node distance=1cm},
    output/.style = {coordinate,node distance=2cm},
    arrow/.style={draw, -latex,node distance=1.5cm},
    pinstyle/.style = {pin edge={latex-, black,node distance=1.5cm}},
    sum/.style = {draw, circle, node distance=1cm}
}

\DeclareUnicodeCharacter{2212}{-} 
\usepackage{pgfplots}
\pgfplotsset{compat=newest}
\pgfplotsset{plot coordinates/math parser=false}
\usepgfplotslibrary{patchplots,groupplots,fillbetween,polar}

\tikzstyle{pinstyle} = [pin edge={to-,thin,black}]

\pgfmathdeclarefunction{gauss}{3}{%
        \pgfmathparse{1/(#3*sqrt(2*pi))*exp(-((#1-#2)^2)/(2*#3^2))}%
     }
   
\usepackage{psfrag}
\usepackage{verbatim}
\usepackage{amsmath}
\usepackage{amsfonts} 
\usepackage{amssymb} 
\usepackage{amsthm}
\usepackage{pifont}
\usepackage{array}
\usepackage{listings}
\usepackage{stfloats}
\usepackage{algorithm} 
\usepackage{algorithmic} 
\usepackage{url} 
\usepackage{enumerate}
\usepackage{multirow}
\usepackage{wasysym}
\usepackage{cancel}
\usepackage{lmodern}
\usepackage{mathrsfs}  
\usepackage{mathtools}
\usepackage{colortbl}
\usepackage{cite}

\input{mydefinitions.tex}

\begin{document}

\title{Clock and Orientation-Robust Simultaneous Radio Localization and Mapping at Millimeter Wave Bands}

  \author{
  \IEEEauthorblockN{Felipe G\'omez-Cuba$^1$, Gonzalo Feijoo-Rodr\'iguez$^1$ and Nuria Gonz\'alez-Prelcic$^2$}
   
\IEEEauthorblockA{
    $^1$atlanTTic, University of Vigo, Spain. Email: \texttt{gomezcuba@gts.uvigo.es,gonfr2000@gmail.com}\\
    $^2$North Carolina State University, USA. Email: \texttt{ngprelcic@ncsu.edu}\\
    }
    
 \thanks{This work has received funding from the Spanish Ministerio de Ciencia e Innovación (MICINN) PID2021-122483OA-I00. This material is based upon work partially supported by the National Science Foundation under Grant 2147955.}
}

\flushbottom
\setlength{\parskip}{0ex plus0.1ex}

\maketitle
\thispagestyle{empty}

\begin{abstract}
This paper proposes a radio simultaneous location and mapping (radio-SLAM) scheme based on sparse multipath channel estimation. By leveraging sparse channel estimation schemes at millimeter wave bands, namely high resolution estimates of the multipath angle of arrival (AoA), time difference of arrival (TDoA), and angle of departure (AoD), we develop a radio-SLAM algorithm that operates without any requirements of clock synchronization, receiver orientation knowledge, multiple anchor points, or two-way protocols. Thanks to the AoD information obtained via compressed sensing (CS) of the channel, the proposed scheme can estimate the receiver clock offset and orientation from a single anchor transmission, achieving sub-meter accuracy in a realistic typical channel simulation.
\end{abstract}

\begin{IEEEkeywords}
Simultaneous location and mapping, compressed sensing, multipath channel estimation, joint localization and channel estimation
\end{IEEEkeywords}

\section{Introduction}

To satisfy the growing demand for broadband, machine-type and low latency applications, wireless standards \cite{3GPPNRoverall16,9090146} incorporate larger bandwidths, large multiple-input multiple-output (MIMO) arrays, and higher frequency bands such as mmWave. Increased sampling rate and directivity allows to resolve the Time Difference of Arrival (TDoA) and Angle of Arrival (AoA) of individual multipath components, respectively. Classic rich scattering models are replaced by Saleh-Valenzuela sparse multipath channel models \cite{Saleh1987}, specially in mmWave frequencies. Since mmWave requires both transmit and receive arrays, sparse recovery algorithms \cite{Venugopal2017,rodriguez2017frequency,8844996,Shahmansoori2018,8761825,palacios2022multidimensional} can estimate not only the receiver TDoA and AoA, but also the transmitter Angle of Departure (AoD). A remarkable by-product is that mmWave sparse channel estimation provides sufficient geometric data to determine the user position without additional location-specific signaling \cite{9356512,Keating2019}. Going one step further, radio simultaneous location and mapping (radio SLAM) of the reflector locations is possible \cite{9473676}.

From the standardization point of view, location methods up to 5G Rel. 16 exploit only the Line-of-Sight (LoS) paths from multiple anchor points of known position \cite{Keating2019}. As in earlier location literature \cite{4103919}, TDoA multilateration requires clock synchronization \cite{7880669,Koivisto2017}, whereas AoA triangulation has to deal with unknown user orientation \cite{Shahmansoori2018,Rastorgueva-Foi2019}. 5G also introduces two-way clock estimation \cite{Keating2019}, which can determine the location combining AoA and TDoA of a single LoS-only anchor \cite{Zhang2018,Abu-Shaban2020}. Sparse multipath NLoS component estimation has spurred new approaches, yet many works inherit methodologies of LoS-only multi-anchor literature, such as treating reflections as “virtual anchor points” \cite{8761910} or assuming a two-way synchronization protocol \cite{Mendrzik2019a}. 

The main contribution of our paper is a simple scheme demonstrating that clock and orientation offset recovery is perfectly viable in \textit{single-anchor one-way} NLoS location. First, we present a linear Least-Squares (LS) location and clock offset recovery algorithm. Building on top of this linear scheme, we recover the user orientation by solving a collection of non-linear single-variable equations.
Prior work in \cite{Wymeersch2018} estimated the orientation assuming two-way clock synchronization. In \cite{Mendrzik2019a} a clock offset is estimated without orientation offset. Several works address clock and/or orientation offset with multiple anchor points \cite{Shahmansoori2018,Rastorgueva-Foi2019,7880669,Koivisto2017} or two-way protocols \cite{Zhang2018,Abu-Shaban2020,Mendrzik2019a}. To the best of our knowledge, our work describes the first scheme that provides both clock and orientation robustness for \textit{single-anchor one-way} multipath radio-SLAM. Moreover, our LS clock recovery scheme is fast enough to be solved many times as part of the orientation recovery algorithm, whereas \cite{Wymeersch2018} relied on a message-passing estimation scheme. A recent pre-print draft presented in \cite{nazari2022mmwave} also proposes clock and orientation recovery relying on iterative message passing.
Finally, we also derive the theoretical Cramer Rao Lower Bound (CRLB) of the receiver and reflector locations. This shows that the CS AoD estimation is not actually necessary for user location nor clock or orientation recovery 
However, the AoDs impact reflector location errors, thus \textit{our analysis shows that CS AoD estimation is a key component of radio-SLAM in single-anchor one-way multipath scenarios}.


The paper is structured as follows: Section \ref{sec:chan} describes the multipath channel geometry and estimation model. Section \ref{sec:loc} describes the clock, orientation and location estimator and the corresponding CRLB. Section \ref{sec:sims} demonstrates the results in simulation. Finally, Section \ref{sec:conclusions} describes conclusions and future work.

%

\section{Multipath Channel Model}
\label{sec:chan}
\begin{figure}[t]
 \centering
 \includegraphics[width=.9\columnwidth]{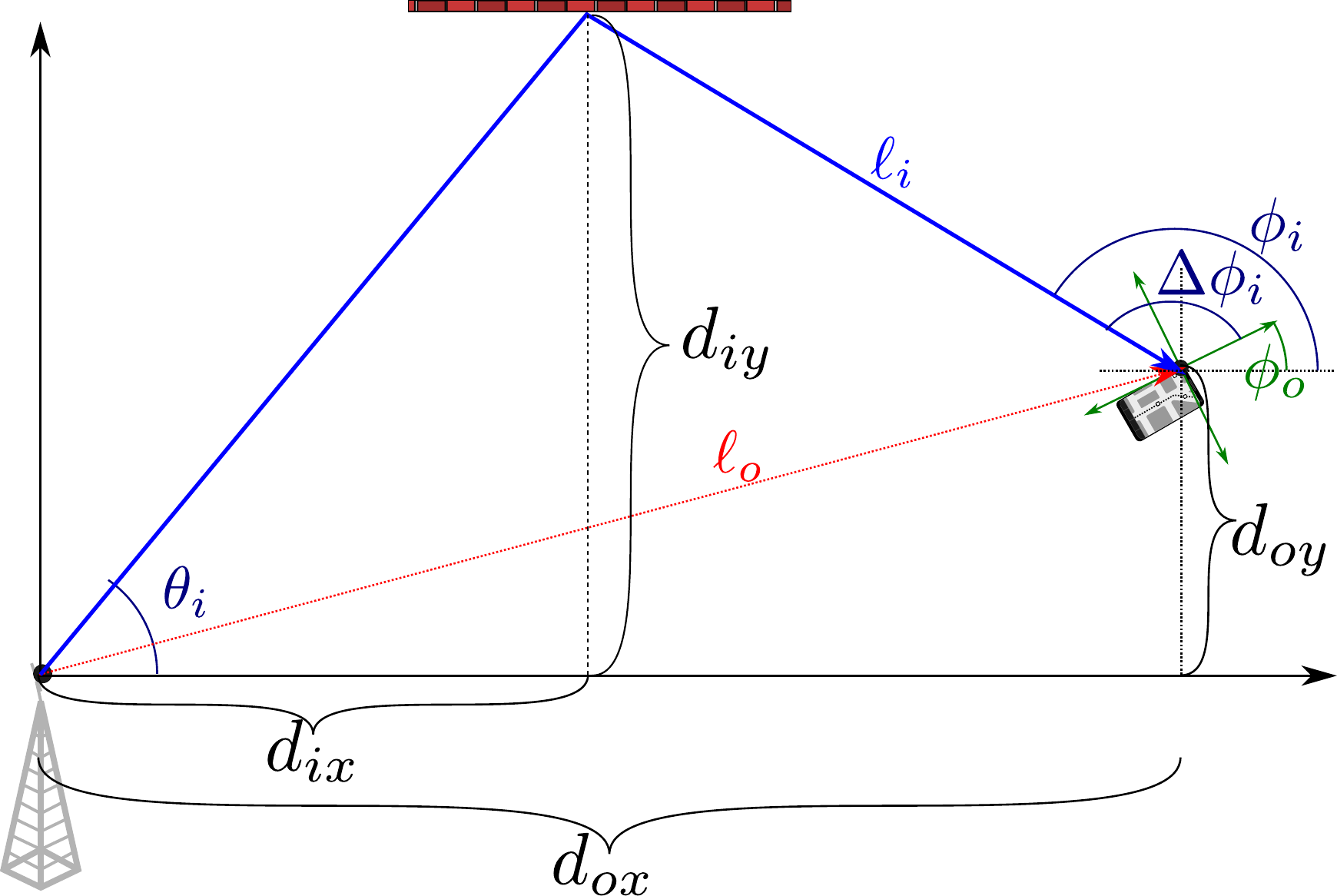}   
 \caption{Geometry of each signal path}
 \label{fig:georay}
\end{figure}

We assume a multipath channel consisting of a set of $N_p$ planar waves that leave a transmitter located at $(0,0)$, and arrive at a receiver located at an unknown position $\dd_o=(d_{ox},d_{oy})$. We assume that only paths with a single reflection, at some unknown position $\dd_i=(d_{ix},d_{iy})$, are strong enough to be detected by the receiver. Fig. \ref{fig:georay} represents the geometric parameters of each path: length equal to the delay multiplied by the speed of light $\ell_i=c\tau_i$, AoD $\theta_i$ and AoA $\phi_i$. Note that, even though this paper presents the 2D case due to space limitations, the results can be extended to 3D. While such extension would be trivial for linear clock offset recovery, the case of 3D orientation can be a bit more nuanced due to the fact that the receiver may experience orientation offset in 3 axes \cite{nazari2022mmwave}.

We assume no clock synchronization. The receiver measures the TDoA of arriving paths $\Delta\tau_i=\tau_i-\tau_e$, from a time reference $\tau_e$ affected by three factors: the direct LoS light travel time $\tau_o=\frac{\sqrt{d_{ox}^2+d_{oy}^2}}{c}$; the synchronization error itself;  and the fact that $\min \tau_i=\tau_o$ for LoS channels, and $\min \tau_i >\tau_o$ for NLoS channels. Similarly, the receiver array is oriented with some unknown rotation $\phi_o$, with paths impinging on the array with difference AoA (DAoA) defined as $\Delta\phi_i=\phi_i-\phi_o$.

Finally, each path suffers attenuation and phase shift characterized by a complex coefficient $\alpha_i$. Though this does not directly affect location, it plays a key role in the performance of sparse multipath channel estimation \cite{8844996}.


\subsection{Sparse Channel Estimation}
\label{sec:cs}
We consider a typical 5G/WiFi7 OFDM waveform for mmWave communication with $N_k$ subcarriers, bandwidth $B$, inter-carrier spacing $\Delta f=B/N_k$, and a cyclic prefix length satisfying $\max \Delta \tau_i \leq T_{CP} \leq \frac{1}{\Delta f}$. The transmitter and receiver are equipped with antenna arrays of size $N_t$ and $N_r$, respectively, connected to $N_{RFt}$ and $N_{RFr}$  RF chains to enable hybrid beamforming (HBF) \cite{Venugopal2017,rodriguez2017frequency}. In addition, the channel is assumed to be invariant for a frame of $N_s$ consecutive OFDM symbols. The equivalent channel at RF port $r$, subcarrier $k$ and symbol $s$ is 
\begin{equation}
\label{eq:ycoefofdm}
y_{s,k,r}=\w_{s,k,r}^H\left(\Hb_{k}\x_{s,k}+\z_{s,k}\right), 
\end{equation}
where $\w_{s,k,r}^H\in\mathbb{C}^{N_{RFr}\times N_r}$ represents the hybrid combiner, while the hybrid precoder is embedded in the pilot $\x_{s,k}$, $\z_{s,k}\sim\mathcal{CN}(0,\sigma_z^2\I_{N_r})$ is i.i.d. additive white gaussian noise (AWGN), and the channel matrix includes the effect of $N_p$ multipath components, i.e. 
\begin{equation}
\label{eq:tchan}
\Hb_{k}=\sum_{i=1}^{N_p} \alpha_i\ab_{N_r}(\Delta\phi_i)\ab_{N_t}(\theta_i)^Te^{-j2\pi k\Delta f\tau_i},
\end{equation}
characterized by array response vectors $\ab_{N_t}(\theta_i)$ and $\ab_{N_r}(\Delta\phi_i)$ vs AoD and DAoA. 

The construction of the CS estimation algorithm is beyond the scope of our paper, but we give here a brief overview. First, we define the vector $\y$ with $y_{s,k,r}$ as its $(sN_kN_{RFr}+kN_{RFr}+r)$-th coefficient.
\begin{equation}
\label{eq:megausian}
\y=\W^H(\I_{N_S} \odot \Hb)\x+\z,
\end{equation}
where $\W=(\w_{0,0,0}|\w_{0,0,1}|\dots|\w_{N_s-1,N_k-1,N_{RFr}-1})$, $\x=(\x_{0,0}^T|\dots|\x_{N_S-1,N_k-1}^T)^T$, $\odot$ is a Kroenecker product, $\Hb$ is a block-diagonal matrix containing $\Hb_0,\dots,\Hb_{N_k-1}$ in its main diagonal,
and $\z$ is an equivalent Gaussian noise vector with elements $z_{s,k,r}=\w_{s,k,r}^H\z_{s,k}$. The noise covariance matrix is also block diagonal $\Sb_{\z}\triangleq\Ex{}{\z\z^H}=\sigma^2_z\texttt{bdiag}\{\W_{s,k}^H\W_{s,k}\}_{s=0,k=0}^{N_S-1,N_k-1}$, with its central elements defined by $\W_{s,k}=(\w_{s,k,0}|\dots|\w_{s,k,N_{RFr}-1})$.



Next, we write $\y$ as a linear combination of columns
\begin{equation}
\label{eq:ysparse}
\y=\sum_{i=1}^{N_p}\alpha_i\vups(\tau_i,\theta_i,\Delta\phi_i)+\z,
\end{equation}
where $\vups(\tau,\theta,\phi)$ is the normalized received signal from a single-path with parameters $(\tau,\theta,\nu)$. We define $\beta^R_{s,k,r}(\phi)=\w_{s,k,r}^H\ab_{n_R}(\phi)$ and $\beta^T_{s,k}(\theta)=\ab_{n_T}(\theta)^T\x_{s,k}$, so that
\begin{equation}
\label{eq:upscoldef}
\upsilon_{s,k,r}(\tau,\theta,\phi) = \beta^R_{s,k,r}(\phi)\beta^T_{s,k}(\theta)e^{-j2\pi k \tau}.
\end{equation}

	Finally, we define a 3D \textit{dictionary} consisting of a set of $K_\tau$ potential TDoAs, $K_\theta$  AoDs and $K_\phi$ DAoAs.
\begin{align}
 \mathcal{D}&=\mathcal{D}_\tau \times \mathcal{D}_\theta \times \mathcal{D}_\phi\\
 \mathcal{D}_\tau&\triangleq \{0,1,\dots,K_\tau-1\} \frac{T_{CP}}{K_\tau}\\
 \mathcal{D}_\theta
 \equiv
 \mathcal{D}_{\Delta\phi}&\triangleq \asin\left(\frac{2}{K_\phi}\left\{-\frac{K_\phi}{2},-\frac{K_\phi}{2}+1,\dots,\frac{K_\phi}{2}-1\right\}\right).
\end{align}
The \textit{observation matrix} is defined with one column per dictionary item $\Upsb_{\mathcal{D}}=(\vups_0|\vups_1|\dots|\vups_{K_\tau K_\theta K_\phi})$.
And the received OFDM signal is finally approximated as
\begin{equation}
\label{eq:ysparse2}
\y \simeq \Upsb_{\mathcal{D}}\s+\z, 
\end{equation}
where $\s$ is a sparse vector with non-zero values $\alpha_i$ in the indices that correspond to the parameters $(\tau_i,\theta_i,\Delta\phi_i)$ for each path of the multipath channel.

Many algorithms can estimate the sparse vector $\s$, implicitly extracting the multipath data. For example, the canonical CS problem subject to a MSE constraint $\xi$ is
$$\min \|\s\|_0 s.t. \|\y-\Upsb_{\mathcal{D}}\s\|_2^2<\xi,$$
a combinatorial problem typically solved using greedy heuristics or $1$-norm relaxation \cite{Venugopal2017,rodriguez2017frequency,8844996}. Moreover, many ML proposals have appeared in recent years \cite{8761825}.
For the rest of the paper we will assume that the location estimator receives an approximation of the real multipath data denoted by $\hat{\mathcal{P}}=\{(\hat{\tau}_i,\hat{\theta}_i,\Delta\hat{\phi}_i)\}_{i=1}^{\hat{N}_p}$, distorted by both dictionary quantization and noise.


\section{Location and Mapping}
\label{sec:loc}
\subsection{Clock-Robust Linear Location}
\label{sec:locknowphi}
We first design a linear scheme to recover $d_{ox}$, $d_{oy}$ and the clock offset $\tau_e$ from the multipath data $\{\Delta\tau_i,\theta_i,\Delta\phi_i\}_{i=1}^{N_{p}}$ assuming the AoA offset $\phi_o$ is known. Therefore, in this subsection we assume that the angles $\phi_i=\Delta \phi_i+\phi_o$ are known. This is reasonable for example in mobile devices with an accelerometer that can measure orientation.

Observing triangles in Fig.\ref{fig:georay}, we combine 
$d_{iy}=d_{ix}\tan\theta_i$ and $d_{iy}-d_{oy}=(d_{ix}-d_{ox})\tan\phi_i,$ to write
%
%
\begin{equation}
\label{eq:triEq}
    d_{ix}\left(\tan\theta_i-\tan\phi_i\right)=d_{oy}-d_{ox}\tan\phi_i.
\end{equation}

Next we compute the length $\ell_i=\ell_e+\Delta\ell_i$ as
\begin{equation}
\label{eq:linEq}
\begin{split}
\ell_e+\Delta\ell_i&=\frac{d_{ix}}{\cos\theta_i}+\frac{d_{ix}-d_{ox}}{\cos\phi_i}\\ 
&=\frac{C_i}{T_i}\left(d_{oy}-d_{ox}\tan\phi_i\right)-\frac{d_{ox}}{\cos\phi_i}\\ 
\end{split}
\end{equation}
where the last steps substitutes $d_{ix}$ from \eqref{eq:triEq}, and defines $T_{i}\triangleq\left(\tan\theta_i-\tan\phi_i\right)$, and $C_i\triangleq\left(\frac{1}{\cos\theta_i}+\frac{1}{\cos\phi_i}\right).$

Next, we define $Q_i\triangleq\frac{C_i}{T_i}=\frac{\cos\phi_i+\cos\theta_i}{\sin\theta_i\cos\phi_i-\cos\theta_i\sin\phi_i}$ and $P_i\triangleq\left(-Q_i\tan\phi_i-\frac{1}{\cos\phi_i}\right)=\frac{\sin\theta_i+\sin\phi_i}{\cos\theta_i\sin\phi_i-\sin\theta_i\cos\phi_i}$, so that
\begin{equation}
\label{eq:linEqtilde}
\begin{split}
d_{ox}P_i+d_{oy}Q_i-\ell_e&=\Delta\ell_i.\\ 
\end{split}
\end{equation}

Since $d_{ox}$, $d_{oy}$ and $\ell_e$ are the same for all paths, we can solve a system of $N_p\geq 3$ linear equations with the Moore-Penrose pseudo-inverse $\B^\dag=(\B^H\B)^{-1}\B^H$ LS solution
\begin{equation}
\label{eq:sislin}
\left(\begin{array}{c}
        \hat{d}_{xo}\\
        \hat{d}_{yo}\\
        \hat{\ell}_e 
     \end{array}\right)=\B^\dag\Delta\boldsymbol{\ell}\textnormal{ with } \B\triangleq\left(\begin{array}{ccc}
        P_1 &Q_1 &-1\\
        P_2 &Q_2 &-1\\
        \vdots&\vdots&\vdots\\
        P_{N_p} &Q_{N_p} &-1\\
     \end{array}\right).
\end{equation}

As is well known, in typical LS estimation problems the error can be reduced by increasing the number of samples. Analogously, it is our intuition that our scheme would perform better when the number of mult-paths $N_p$ increases. On the other hand, the size of the dictionaries and sparsity of the channel has a sophisticated impact on CS channel estimation performance \cite{8844996}. We leave for future work a more systematic analytical characterization of the influence of the parameters $N_p$, $K_\tau$, $K_\theta$, $K_\phi$, and $\{\alpha_i\}_{i=1}^{N_p}$.

Finally, the LoS distance and delay follow from
$\hat{\ell}_o=c\hat{\tau}_o=\sqrt{\hat{d}_{xo}^2+\hat{d}_{yo}^2}$. The reflector locations from \eqref{eq:triEq} and $\hat{d}_{yi}=\hat{d}_{xi}\tan\theta_i$. The clock offset is $\frac{\min \hat{\ell_i}}{c}-\min\Delta\tau_i$, and the LoS condition follows by comparing $\hat{\ell}_o$ vs $ \min\hat{\ell}_i$.

\subsection{Orientation Recovery}
 \label{sec:orec}
Next, we assume $\phi_o$ is unknown. Therefore, substituting $\phi_i=\Delta \phi_i+\phi_o$ into \eqref{eq:sislin} we define the following functions
$$P_i(\phi_o)\triangleq\frac{\sin\theta_i+\sin(\Delta\phi_i+\phi_o)}{\cos\theta_i\sin(\Delta\phi_i+\phi_o)-\sin\theta_i\cos(\Delta\phi_i+\phi_o)},$$
$$Q_i(\phi_o)\triangleq\frac{\cos(\Delta\phi_i+\phi_o)+\cos\theta_i}{\sin\theta_i\cos(\Delta\phi_i+\phi_o)-\cos\theta_i\sin(\Delta\phi_i+\phi_o)}.$$
To determine $\phi_o$, we propose the following method: 
\begin{enumerate}
 \item Divide the set $\{1\dots N_p\}$ into $N_G$ groups of paths 
 $$\mathcal{G}_1,\mathcal{G}_2\dots\mathcal{G}_{N_G}$$
 \item Construct $N_G$ separate systems of linear equations \eqref{eq:sislin} for each group $\mathcal{G}_1,\mathcal{G}_2\dots\mathcal{G}_{N_G}$ and obtain the solutions of each as a function of $\phi_o$
\begin{equation}
\label{eq:Fmdef}
F_m(\phi_o)=\left(\begin{array}{c}
        \hat{x}_o^{(m)}\\
        \hat{y}_o^{(m)}\\
        \tilde{\hat\ell}_o^{(m)} 
     \end{array}\right)=\B_{\mathcal{G}_m}(\phi_o)^\dag\Delta\boldsymbol{\ell}_{\mathcal{G}_m}
\end{equation}
\item For the correct value of $\phi_o$, all groups should calculate the same location. Therefore, we can simply solve the system of non-linear equations of $\phi_o$
$$F_{1}(\phi_o)=F_{2}(\phi_o)=\dots=F_{N_G}(\phi_o).$$
However, in order to support error in the parameters $\{(\hat{\tau}_i,\hat{\theta}_i,\hat{\phi}_i)\}_{i=1}^{\hat{N}_p}$, the system of equations is solved \textbf{approximately} in the sense of minimizing the mean squared distance between the computed locations
\begin{equation}
\label{eq:minmselin}
 \min_{\phi_o'}\sum_{m=1}^{N_G} \left\|F_{m}(\phi_o')-\overline{F}(\phi_o')\right\|^2\textnormal{ s.t. }\overline{F}(\phi_o')=\frac{\sum_{m=1}^{N_G} F_{m}(\phi_o')}{N_G}
\end{equation}
\end{enumerate}

\begin{figure}[t]
 \centering
 \subfigure[3P groups location]{
    \includegraphics[trim= {0  0.25cm  0 1.3cm},clip,width=.9\columnwidth]{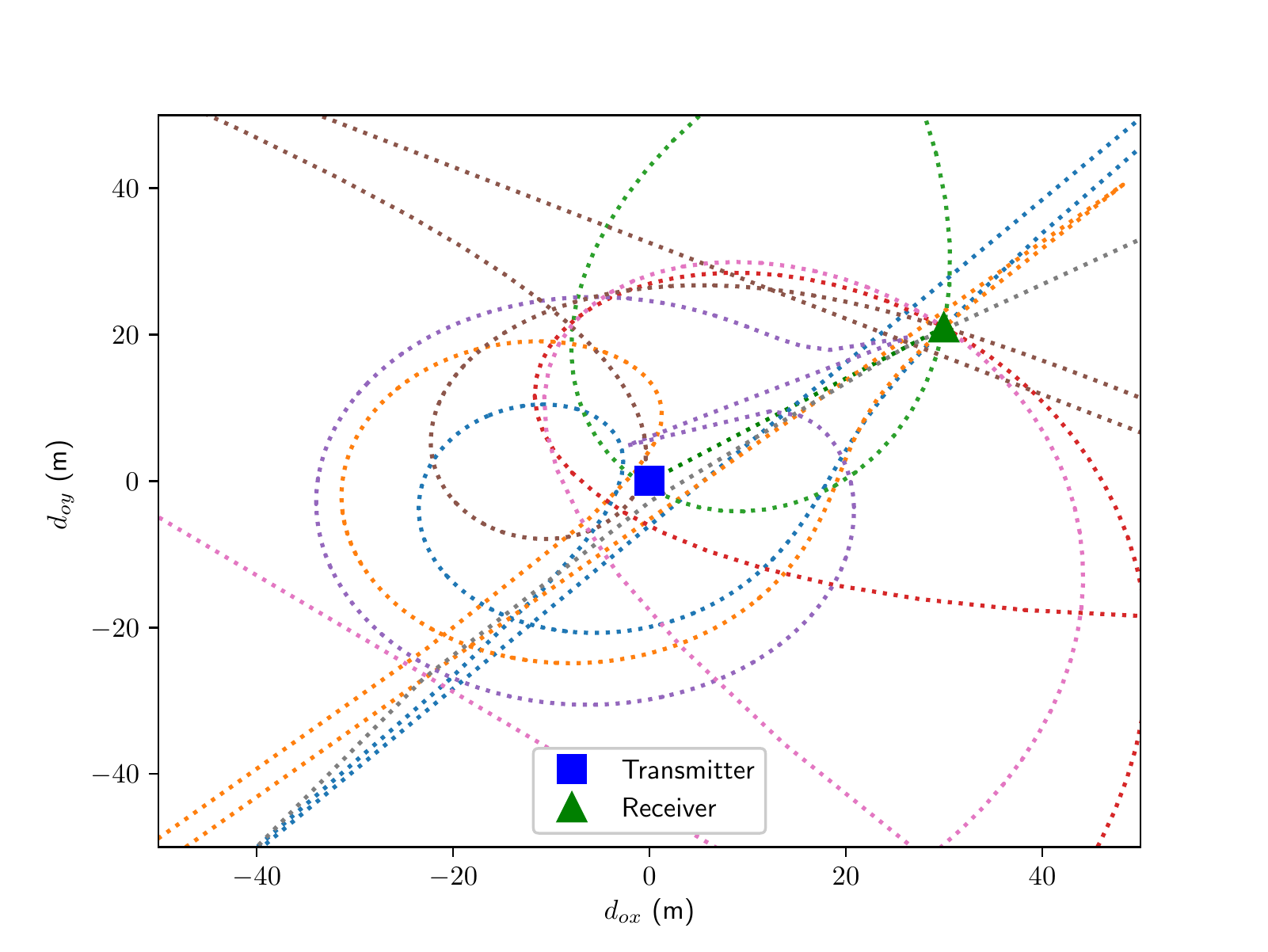}
    \label{fig:graphsol3P}
 }
 \subfigure[D1 groups location]{
 \includegraphics[trim= {0  0.25cm  0 1.3cm},clip,width=.9\columnwidth]{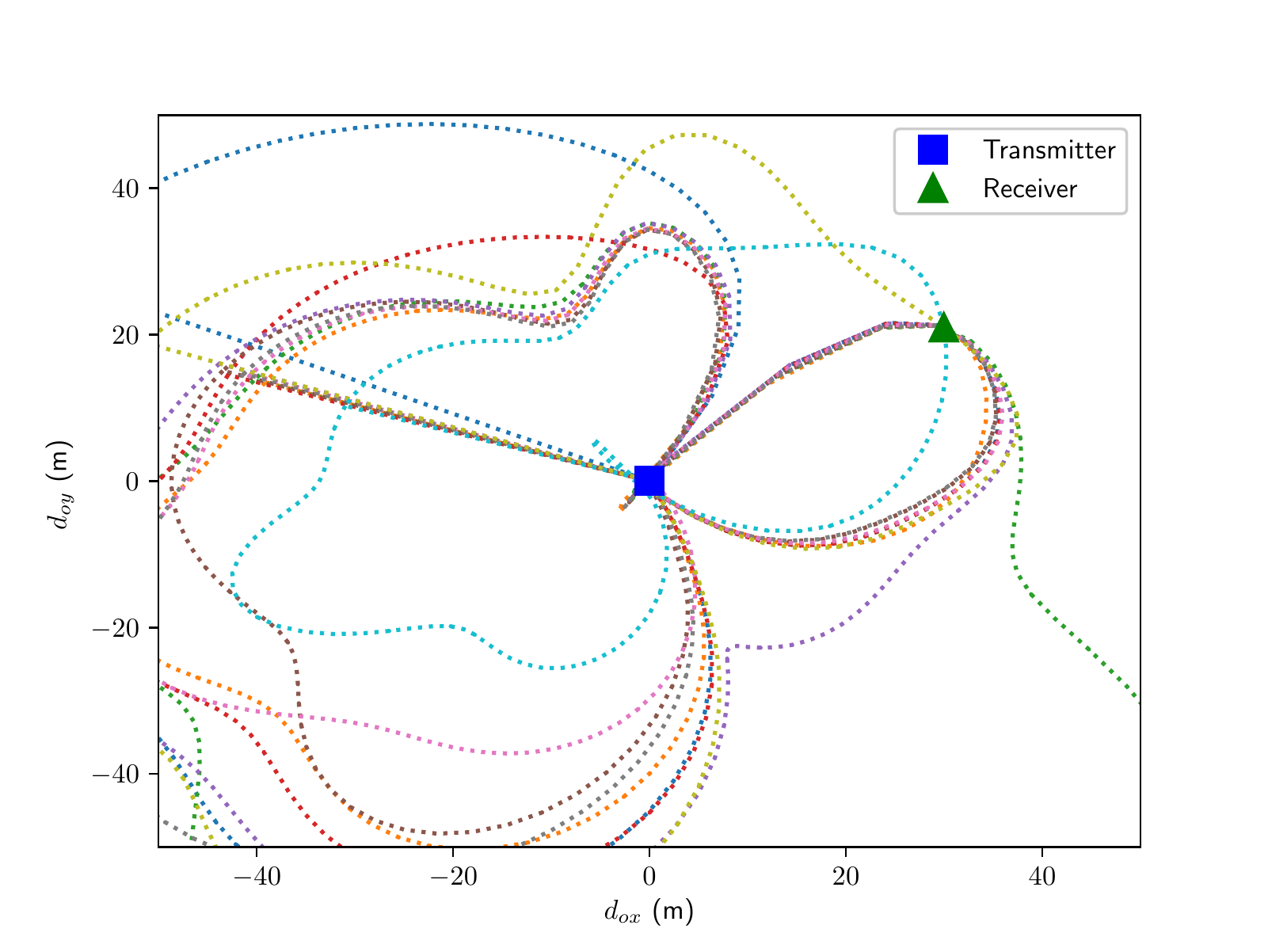}
    \label{fig:graphsolD1}
 }
 \caption{Estimated $\hat{d}_{ox}$ and $\hat{d}_{oy}$ with $N_p=10$. Each represented trajectory depicts one group's location guess vs different values of $\phi_o'\in[0,2\pi)$. At the point where all trajectories meet, the groups are in agreement, and the correct value of $\phi_o$ is found.}
 \label{fig:graphsolboth}
\end{figure}

Our scheme admits arbitrary grouping strategies. For example, a sliding window of ``three path'' (3P) groups $\mathcal{G}_1={1,2,3}$, $\mathcal{G}_2={2,3,4}$, $\dots$ $\mathcal{G}_m={m,m+1,m+2}$ (the minimum necessary for the linear system). Or the ``drop one'' (D1) collection of complementary groups containing ``all paths except the $m$-th'', $\mathcal{G}_m=\{1\dots m-1,m+1\dots N_p\}$. We compare these two strategies in Fig. \ref{fig:graphsolboth}. Different grouping strategies present different solution shapes and intersections, thus the design of optimal grouping strategies is left for future work. We note that $N_G\geq3$ should be used to discard points where only two lines cross.

\subsection{Location Error Bounds}
  
Denoting the multipath data vector $\m=(\m_1^T,\m_2^T,\dots,\m_{N_p}^T)^T$, with $\m_{i}=(\theta_i,\phi_i,\tau_i)^T$. In any unbiased classic (not Bayesian) estimator $\hat{\m}$ of $\m$, the error covariance Cramer Rao Lower Bound \cite{Shahmansoori2018} is
  $$\Sb_{\m}\triangleq\Ex{}{(\m-\hat{\m})(\m-\hat{\m})^T}\succeq \J_{\m}^{-1}$$
  where the Fisher Information Matrix (FIM) is defined as
  \begin{equation}
  \label{eq:defisher}
   \J_{\m}=\Ex{\y|\m}{-\frac{\partial^2 \log \mathscr{L}_\y(\m) }{\partial\m\partial\m^T}}
  \end{equation}

 The Gaussian vector $\y|\m$ in \eqref{eq:megausian} has mean $\vmu_{\y}=\W^H(\I_{n_S} \odot \Hb)\x$ and covariance $\Sb_{\z}$.
We define $\R$ such that $\Sb_{\z}^{-1}=\frac{\R^H\R}{\sigma_z^2}$ and generalize the proof in \cite{Shahmansoori2018} to obtain
  \begin{align*}
    \J_{\m}^{-1}
        &=\frac{\sigma_z^2}{2}\Re\left\{ \frac{\partial \vmu_{\y}^H}{\partial \m }\R^H\R\frac{\partial \vmu_{\y}}{\partial \m^T }\right\}^{-1}
  \end{align*}
where, recalling \eqref{eq:ysparse}-\eqref{eq:upscoldef}, each term satisfies
$$\frac{\partial \mu_{y_{s,k,r}}}{\partial \theta_i }=\alpha_ie^{-j2\pi k \tau_i}\beta^R_{s,k,r}(\Delta\phi_i)\frac{\partial \beta^T_{s,k}(\theta_i)}{\partial \theta_i }$$
$$\frac{\partial \mu_{y_{s,k,r}}}{\partial \Delta\phi_i }=\alpha_i\beta^T_{s,k}(\theta_i)e^{-j2\pi k \tau_i}\frac{\partial \beta^R_{s,k,r}(\Delta\phi_i)}{\partial \Delta\phi_i }$$
$$\frac{\partial \mu_{y_{s,k,r}}}{\partial \tau_i }=-j2\pi k\alpha_i\beta^R_{s,k,r}(\Delta\phi_i)\beta^T_{s,k}(\theta_i)e^{-j2\pi k \tau_i}$$
  
To transform the above into a covariance of \textbf{location} error, we define $\dd=(\dd_o,\tau_e,\phi_o,\dd_1,\dots,\dd_{N_p})$, and write
\begin{equation}
\label{eq:crlbloc}
\Sb_{\dd}\triangleq \Ex{}{(\dd-\hat{\dd})(\dd-\hat{\dd})^T}\succeq \J_{\dd}^{-1} = (\T^T\J_{\m}\T)^{-1}
\end{equation}
where the transformation matrix is 
$\T^T=\frac{\partial \m^T}{\partial \dd}$, that is, the derivatives vs $\dd_o$, $\tau_e$, $\phi_o$, and $\dd_i$ of the expression:

%
\begin{align*}
 \Delta \tau_i &= \frac{\|\dd_i-\dd_o\|+\|\dd_i\|}{c} - \tau_e\\
 \theta_i &= \arctan\left(\frac{d_{iy}}{d_{ix}}\right)+\pi\mathbb{I}_{d_{ix}<0}\\
 \Delta \phi_i &= \arctan\left(\frac{d_{iy}-d_{oy}}{d_{ix}-d_{ox}}\right)+\pi\mathbb{I}_{d_{ix}<d_{ox}} - \phi_o.
\end{align*}
The derivatives are given in Table \ref{tab:derivs}. By inspection, the structure of $\T^T$ features two fully non-zero top rows, $\T_o^T=\left(\frac{\partial \m_1^T}{\partial \dd_{o}},\frac{\partial \m_2^T}{\partial \dd_{o}},\dots,\frac{\partial \m_{N_p}^T}{\partial \dd_{o}}\right)$; the next two rows are of the form $(-1,0,0,-1\dots)$, representing the mutual dependence of all paths on the clock and orientation, and finally the last $2N_p$ rows of $\T^T$ constitute a block-diagonal submatrix with elements $\T_i^T=\frac{\partial \m_i^T}{\partial \dd_{i}}$.

\begin{table}[b]
\centering
\caption{Derivatives forming the matrix $\T^T$}
\label{tab:derivs}
\begin{tabular}{c|ccc}
    &$\partial \tau_i$&$\partial \theta_i$&$\partial \phi_i$\\\hline
$\frac{1}{\partial \dd_o}$&$\frac{\dd_o}{c\|\dd_i-\dd_o\|}$&$0$&$\frac{(d_{iy}-d_{oy},d_{ox}-d_{ix})}{\|\dd_i-\dd_o\|^2}$\\
$ \frac{1}{\partial \tau_o}$ &$-1$&$0$&$0$\\
$ \frac{1}{\partial \phi_o}$ &$0$&$0$&$-1$\\
$ \frac{1}{\partial \dd_i}$ &$\frac{\dd_i}{c\|\dd_i-\dd_o\|}+\frac{\dd_i}{c\|\dd_i\|}$&$\frac{(-d_{iy},d_{ix})}{\|\dd_i\|^2}$&$\frac{(d_{oy}-d_{iy},d_{ix}-d_{ox})}{\|\dd_i-\dd_o\|^2}$\\
\end{tabular}
\end{table}

We may distinguish the error of \textit{user location}, $\J_{\dd_o}=\T_o^T\J_{\m}\T_o$, as the first $2\times 2$ submatrix of $\J_{\dd}$, where the derivative $ \frac{\partial{\theta_i}}{\partial \dd_o} = 0$ suggests that the AoDs estimation are not necessary in to estimate the user location error, $\tau_e$ and $\phi_o$. Indeed, location algorithms employing AoA and TDoA only do exist. On the other hand, the error of \textit{reflector mapping} $\J_{\dd_1,\dots,\dd_{N_p}}$, does indeed depend on the AoDs. In other words, our theoretical error analysis shows that AoD information as a by-product of CS channel estimation is a key enabler for single-anchor one-way radio-SLAM.


\section{Simulation Results}
\label{sec:sims}
We simulate\footnote{The simulation files are openly available at \texttt{https://github.com/gomezcuba/CASTRO-5G}} $N_ {sim}=1000$ user locations $(d_{ox},d_{oy})$, and $N_{p}\times N_{sim}=20\times 1000$ reflector locations $(d_{ix},d_{iy})$. Both receiver and reflector locations are generated with a uniform random distribution in a square of size 100x100m. The terminal orientation $\phi_o$ is also random $U(0,2\pi)$ and the clock offset $\tau_e-\tau_o$ is distributed as $U(0,40ns)$. 

\subsection{Perfect multipath information}
\begin{figure}
 \centering
 \subfigure[Perfect multipath information]{
    \includegraphics[trim= {0  0.25cm  0 1.3cm},clip,width=.9\columnwidth]{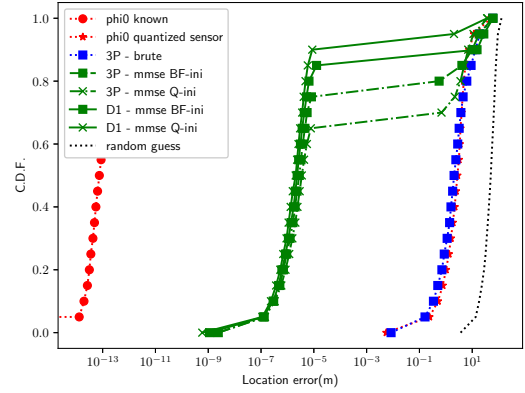}
    \label{fig:locerrcdf}
    }
 \subfigure[Multipath info. $K_{\phi}=256$]{
    \includegraphics[trim= {0  0.25cm  0 1.3cm},clip,width=.9\columnwidth]{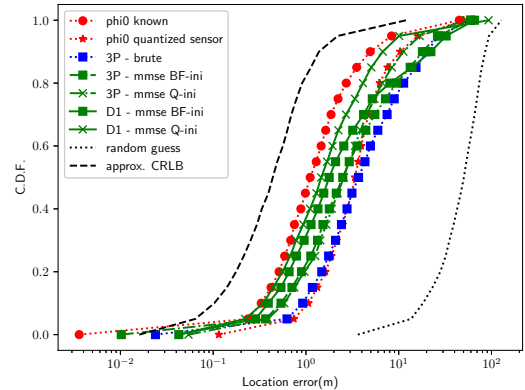}
 \label{fig:locerrcdfDic} 
    }
 \caption{Location error CDF}
\end{figure}

The fist simulation confirms the robust location method with unknown $\tau_e$ and $\phi_o$ with perfectly known multipath information.
Fig. \ref{fig:locerrcdf} presents location error C.D.F. in meters ($\sqrt{(p_{ox}-\hat{p}_{ox})^2+(p_{oy}-\hat{p}_{oy})^2}$). We depict the solutions to the system of linear equations \eqref{eq:sislin} in dotted red lines of two types: with round markers, we represent the solution of \eqref{eq:sislin} when $\phi_o$ is perfectly known. On the other hand, with star markers we present a more realistic mobile phone equiped with accelerometers, in which a quantized orientation $\mathcal{Q}_{N_Q}(\phi_o)=\arg\min_{n\in\{0,N_Q-1\}}|\phi_o-n\frac{2\pi}{N_Q}|$ is known (we use $N_Q=64$). This shows that with known $\phi_o$ the clock error is almost perfectly corrected ($10^{-12}$m error is nearly the floating point precision). A coarse orientation knowledge $\mathcal{Q}_{N_Q}(\phi_o)$ leads to 10 orders of magnitude degradation of the location error ($10$cm-$1m$). For a better intuition, we depict in dotted black a ``random guess'' worst-case estimator that generates $(\hat{d}_{ox},\hat{d}_{oy})$ as an independent uniform distribution. This shows that even in the case of poor orientation knowledge from a gyroscope, a significant improvement on the information on the user location can be extracted from the multipath data (compared to the initial maximum uncertainty).

We explore different estimations of $\hat{\phi_o}$ by solving the system of non-linear equations approximately in a MMSE sense \eqref{eq:minmselin}, employing different algorithms and adopting different ``grouping schemes'' as discussed in Subsection \ref{sec:orec}. The 3P scheme (dashed line) performs $(N_p-2)$ inversions of $3\times 3$ matrices, and is much faster than the D1 scheme (solid line), which needs to invert $N_p$ $(N_p-1)\times(N_p-1)$ matrices. In the faster 3P method, a brute-force grid search of the minimum \eqref{eq:minmselin} in the interval \texttt{linspace(0,2$\pi$,100)} is factible. This brute force search is nearly as coarse as the quantizer, and achieves similar location error.

Otherwise, we employ the built-in Python library \texttt{scipy.optimize.root()} with \texttt{method='ML'} to solve \eqref{eq:minmselin}. This tool requires an initialization point, where we compare initialization using the 3P brute force result, versus initialization based on the assumption that the mobile has an orientation sensor that coarsely estimates a quantized orientation $\mathcal{Q}_{N_Q}(\phi_o)$, as discussed in the previous paragraph. For all algorithms, the location error is $1-10$ $\mu$m in most realizations, suggesting that $\phi_o$ is recovered with much more precision than the quantizer/brute-force search. The 3P scheme using $\mathcal{Q}_{N_Q}(\phi_o)$ as initialization seems to be stuck at the initialization point the most among all algorithms, for 30\% of realizations; this can be explained by the extreme asymptotic lines observed in Fig. \ref{fig:graphsol3P}, which suggest that the 3P grouping scheme leads to a poor conditioning of the non convex optimization \eqref{eq:minmselin}.

Overall, simulations show that clock and orientation offsets can be compensated achieving near-perfect location when the multipath information is perfect. Of course, location precisions of the order of $\mu$m are not truly achievable in practical systems where the multipath parameters $\{\Delta\tau_i,\theta_i,\Delta\phi_i\}_{i=1}^{N_{p}}$ are estimated with some error. This result must be interpreted to mean that there are no intrinsic physical limitations to how well clock and orientation error can be corrected if $\{\Delta\tau_i,\theta_i,\Delta\phi_i\}_{i=1}^{N_{p}}$ are known perfectly. The linear clock compensation is straightforward, whereas in the non-linear orientation compensation \eqref{eq:minmselin} the root algorithm is less important than the ``group selection'' method, which plays a pivotal role in the trade-off between complexity and accuracy.

\subsection{Multipath Information with Dictionary Quantization}

We now consider the recovery of the multipath components with error due to using a discrete dictionary. 
Since the design of practical CS algorithms is beyond the scope of our paper, we consider that the multipath recovery error is modelled as a perfect quantization of the true multipath information, rather than noise. We shall explore the impact of different multipath parameters $\theta_i$, $\phi_i$ and $\tau_i$ separately. We focus on the case of imperfect DAoAs, where the location algoritm receives $\mathcal{Q}_{K_{\phi}}(\Delta\phi_i)$ where $K_{\phi}\geq N_{r}$ is the dictionary size. We remark that even though typical mmWave uniform linear arrays (ULA) and uniform planar arrays (UPA) have 10-100 antennas, the angular dictionary can implement CS \textit{superresolution} \cite{Venugopal2017,rodriguez2017frequency,8844996} and be much larger. In the first experiment we test the impact of AoA quantization only, assuming the  AoDs $\theta_i$ and TDoA $\Delta \tau_i$ are perfectly known. 

As an approximation of the CRLB of location error, we consider that the DAoA errors follow an uniform distribution $\mathcal{Q}_{K_{\phi}}(\Delta\phi_i)-\Delta\phi_i\sim U(-\frac{ \pi}{2K_{\phi}},-\frac{ \pi}{2K_{\phi}})$. Therefore $\Sb_{\Delta \boldsymbol{\phi}}= \frac{1}{12}\left(\frac{\pi}{K_{\phi}}\right)^2\I_{20}$. We replace $\J_{\m}$ in \eqref{eq:crlbloc} by this variance matrix, in combination with table \ref{tab:derivs}, to calculate a semi-numerical CRLB approximation:

$$\textnormal{approxCRLB}=\sqrt{\frac{1}{12}\left(\frac{\pi}{K_{\phi}}\right)^2\tr\{(\T_o^T\T_o)^{-1}\}}$$

Fig. \ref{fig:locerrcdfDic} represents the location error CDF with the same algorithms in the previous section, but with AoD quantization $K_{\theta}=256$. Despite the fact that the quantization is less than $0.4\%$ of the circumference, the location accuracy is severely degraded. Due to imprecise multipath information, now the location error only stays $<2.5m$ for $80\%$ of the realizations even with perfectly known $\phi_o$ and $<4.5$m with a quantized orientation sensor $\mathcal{Q}_{N_Q}(\phi_o)$. These results are reasonably close to the CRLB, which is $<1m$ for the $80\%$-ile. The best case of estimated $\hat{\phi_o}$ is the D1 grouping scheme with $\mathcal{Q}_{N_Q}(\phi_o)$ as initialization, which performs very close to the case of known $\phi_o$. The D1 grouping scheme using the 3P brute-force result as initialization also performs reasonably, as it outperforms the 3P grouping schemes with either brute-force or \texttt{root()} solvers with any initialization. Again we observe that the choice of the grouping scheme in Subsection \ref{sec:orec} is critical, as the 3P methods perform even worse than a raw quantized orientation sensor in $30\%$ of the realizations. We remark that in this paper we have yet not introduced ``multipath refinement'' algorithms \cite{Shahmansoori2018} nor message-passing interactions between multipath, position and clock offset recovery \cite{Wymeersch2018}. Thus, we show that \textit{with very simple algorithms, one-way single-anchor multipath location can estimate user locations up to few-meters precision}. This is useful as-is, though the integration of our orientation and clock recovery schemes with iterative refinement estimation is a promising improvement for future work.

\begin{figure}
 \centering
 \subfigure[Receiver Location Error]{
    \includegraphics[trim= {0 0.25cm 0 1.3cm},clip,width=.9\columnwidth]{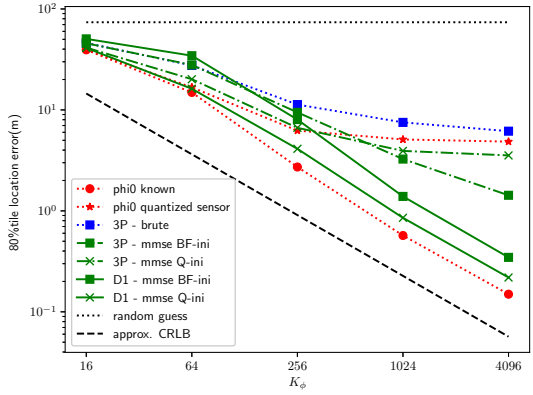}
 }
 \subfigure[Reflector Location Error]{
    \includegraphics[trim= {0 0.25cm 0 1.3cm},clip,width=.9\columnwidth]{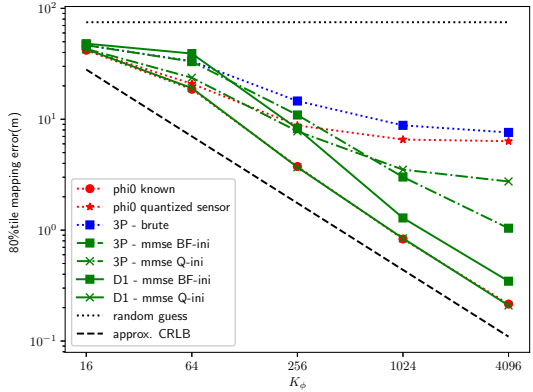}
    }
 \subfigure[Clock Offset Error]{
    \includegraphics[trim= {0  0.25cm  0 1.3cm},clip,width=.9\columnwidth]{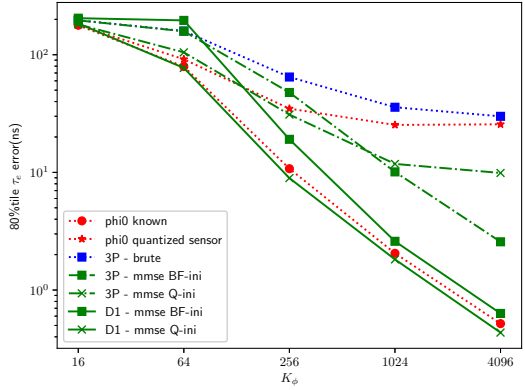}
    \label{fig:Kdvstaue}
    }
 \subfigure[Orientation Offset Error]{
    \includegraphics[trim= {0  0.25cm  0 1.3cm},clip,width=.9\columnwidth]{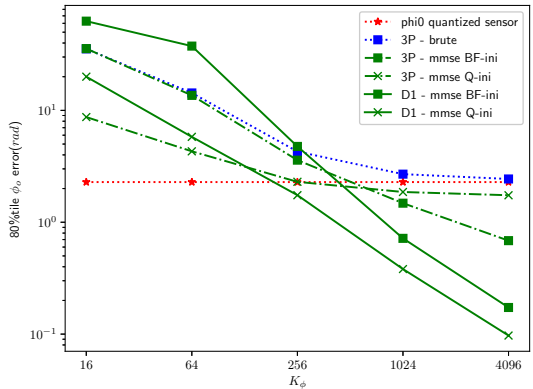}
    \label{fig:Ktvsphi0e}
    }
 \caption{Radio SLAM error $80\%$-ile vs $K_\phi$.}
    \label{fig:Ktvserror}
\end{figure}

Having established the behavior of the estimators, we focus on the $80\%$-iles of location error vs $K_{\phi}$, depicted in Fig.  \ref{fig:Ktvserror}. While the inherent quantization of either gyroscope orientation ($\mathcal{Q}_{N_Q}(\phi_o)$) or brute-force search created error floors, our orientation estimation scheme with D1 grouping can almost achieve the same performance as the case with known $\phi_o$. Moreover, the reflector and receiver location results are very similar. This occurred throughout all our simulations. 
Likewise, we performed simulations with quantized DAoA and TDoA which resulted in very similar plots, which are omitted due to the page constraints.
Finally, Fig. \ref{fig:Kdvstaue} shows the clock offset estimation error vs $K_{\phi}$, showing that clock offset estimation within a fraction of a nano-second is feasible using only one-way single-anchor multipath data obtained from a practical dictionary-based CS channel estimation scheme. Likewise, in Fig. \ref{fig:Ktvsphi0e} the orientation error achieves $0.1$ radian accuracy.

\section{Conclusions}
\label{sec:conclusions}

We have derived a simple and robust \textit{single-anchor one-way} multipath radio-SLAM scheme with clock and orientation offset recovery. We have demonstrated that multiple-anchor and two-way protocols typical in LOS-only location techniques are not required for clock and orientation robust location.
The proposed algorithm employs a simple LS method for location and clock recovery, and a single-variable non-linear equation root method for orientation estimation. Moreover, we have developed a CRLB error analysis that takes into account the HBF precoding matrices. Even with a low complexity implementation, we have have
seen that location accuracies of about $10$cm-$1m$, clock accuracies of $0.1$ns and orientation accuracies of $0.1^o$  are achievable.


\end{document}

%% file: mydefinitions.tex
\DeclareMathOperator{\asin}{asin}

\newcommand{\Hb}{\mathbf{H}}
\newcommand{\Sb}{\mathbf{\Sigma}}

\newcommand{\B}{\mathbf{B}}

\newcommand{\T}{\mathbf{T}}
\newcommand{\I}{\mathbf{I}}

\newcommand{\W}{\mathbf{W}}

\newcommand{\R}{\mathbf{R}}

\newcommand{\Upsb}{\mathbf{\boldsymbol{\Upsilon}}}

\newcommand{\J}{\mathbf{J}}

\newcommand{\ab}{\mathbf{a}}

\newcommand{\dd}{\mathbf{d}}

\newcommand{\m}{\mathbf{m}}

\newcommand{\s}{\mathbf{s}}

\newcommand{\w}{\mathbf{w}}
\newcommand{\x}{\mathbf{x}}
\newcommand{\y}{\mathbf{y}}
\newcommand{\z}{\mathbf{z}}

\newcommand{\vmu}{\mathbf{\boldsymbol{\mu}}}

\newcommand{\vups}{\mathbf{\boldsymbol{\upsilon}}}

\newcommand{\tr}{\textnormal{tr}}

\newcommand{\Ex}[2]{{\mathbb{E}_{#1}\left[#2\right]}}

\newtheorem{definition}{Definition}
\newtheorem{lemma}{Lemma}
\newtheorem{corollary}{Corollary}

\theoremstyle{plain}

\newtheorem{example}{Example}

   \definecolor{blueH3}{rgb}{0,.5,1}
   \definecolor{blueH2}{rgb}{0,0.25,0.75}
   \definecolor{blueH1}{rgb}{0,0,0.5}   
   \definecolor{grayOldText}{rgb}{.5,.5,.5}
   \definecolor{VCobalt}{HTML}{005682}
   \definecolor{TZTeal}{HTML}{008080}
   \definecolor{TZTealfaded}{HTML}{F0FFFF}
   \definecolor{KYJade}{HTML}{008151}
   \definecolor{ARust}{HTML}{a10000}
   \definecolor{FFucsia}{HTML}{7000c3}   
   \definecolor{Tangerine}{HTML}{d45500}
   
\newcommand{\CASE}[1]{\STATE \textbf{case} #1\textbf{:} \begin{ALC@g}}
\newcommand{\ENDCASE}{\end{ALC@g}}

\newcommand{\DEFAULT}{\STATE \textbf{default:} \begin{ALC@g}}
\newcommand{\ENDDEFAULT}{\end{ALC@g}}
\newcommand{\DEFAULTLINE}[1]{\STATE \textbf{default:} }

\makeatletter
\newcommand\remembertext[2]{
  \immediate\write\@auxout{\unexpanded{\global\long\@namedef{mytext@#1}{#2}}}%
  #2%
}
\newcommand\recalltext[1]{%
  \ifcsname mytext@#1\endcsname
    \@nameuse{mytext@#1}%
  \else
    ``??''
  \fi
}

\newcounter{rcnt}
\newcounter{ccnt}

\newlength{\ansspace}
\newlength{\stdleftskip}
\newlength{\stdrightskip}
\newlength{\citeskip}

%% file: mplocwcnc.bbl
\begin{thebibliography}{10}
\providecommand{\url}[1]{#1}
\csname url@samestyle\endcsname
\providecommand{\newblock}{\relax}
\providecommand{\bibinfo}[2]{#2}
\providecommand{\BIBentrySTDinterwordspacing}{\spaceskip=0pt\relax}
\providecommand{\BIBentryALTinterwordstretchfactor}{4}
\providecommand{\BIBentryALTinterwordspacing}{\spaceskip=\fontdimen2\font plus
\BIBentryALTinterwordstretchfactor\fontdimen3\font minus
  \fontdimen4\font\relax}
\providecommand{\BIBforeignlanguage}[2]{{%
\expandafter\ifx\csname l@#1\endcsname\relax
\typeout{** WARNING: IEEEtran.bst: No hyphenation pattern has been}%
\typeout{** loaded for the language `#1'. Using the pattern for}%
\typeout{** the default language instead.}%
\else
\language=\csname l@#1\endcsname
\fi
#2}}
\providecommand{\BIBdecl}{\relax}
\BIBdecl

\bibitem{3GPPNRoverall16}
3GPP, ``{3rd Generation Partnership Project; Technical Specification Group
  Radio Access Network; NR; NR and NG-RAN Overall Description; Stage 2 (Release
  16)},'' \emph{ETSI TS 38.300}, no. 16.5.0, 2021.

\bibitem{9090146}
E.~Khorov, I.~Levitsky, and I.~F. Akyildiz, ``{Current Status and Directions of
  IEEE 802.11be, the Future Wi-Fi 7},'' \emph{IEEE Access}, vol.~8, pp.
  88\,664--88\,688, 2020.

\bibitem{Saleh1987}
A.~A.~M. Saleh and R.~Valenzuela, ``{A statistical model for indoor multipath
  propagation},'' \emph{IEEE Journal on Selected Areas in Communications},
  vol.~5, no.~2, pp. 128--137, feb 1987.

\bibitem{Venugopal2017}
K.~Venugopal, A.~Alkhateeb, N.~G. Prelcic, and R.~W. Heath, ``{Channel
  estimation for hybrid architecture based wideband millimeter wave systems},''
  \emph{IEEE Journal on Selected Areas in Communications}, vol.~35, no.~9, pp.
  1996--2009, jun 2017.

\bibitem{rodriguez2017frequency}
J.~Rodriguez-Fernandez, N.~{Gonz{\'{a}}lez Prelcic}, K.~Venugopal, and R.~W.
  {Heath Jr.}, ``{Frequency-domain compressive channel estimation for
  frequency-selective hybrid mmWave MIMO systems},'' \emph{IEEE Transactions on
  Wireless Communications}, vol.~17, no.~5, pp. 2946--2960, mar 2018.

\bibitem{8844996}
F.~G{\'{o}}mez-Cuba and A.~J. Goldsmith, ``{Compressed sensing channel
  estimation for OFDM with non-Gaussian multipath gains},'' \emph{IEEE
  Transactions on Wireless Communications}, vol.~19, no.~1, pp. 47--61, 2020.

\bibitem{Shahmansoori2018}
A.~Shahmansoori, G.~E. Garcia, G.~Destino, G.~Seco-Granados, and H.~Wymeersch,
  ``{Position and Orientation Estimation Through Millimeter-Wave MIMO in 5G
  Systems},'' \emph{IEEE Transactions on Wireless Communications}, vol.~17,
  no.~3, pp. 1822--1835, 2018.

\bibitem{8761825}
F.~Zhu, A.~Liu, and V.~K.~N. Lau, ``{Channel Estimation and Localization for
  mmWave Systems: A Sparse Bayesian Learning Approach},'' in \emph{ICC 2019 -
  2019 IEEE International Conference on Communications (ICC)}, 2019, pp. 1--6.

\bibitem{palacios2022multidimensional}
\BIBentryALTinterwordspacing
J.~Palacios, N.~Gonz{\'{a}}lez-Prelcic, and C.~Rusu, ``{Multidimensional
  orthogonal matching pursuit: theory and application to high accuracy joint
  localization and communication at mmWave},'' \emph{arXiv preprint}, 2022.
  [Online]. Available: \url{https://arxiv.org/abs/2208.11600}
\BIBentrySTDinterwordspacing

\bibitem{9356512}
O.~Kanhere and T.~S. Rappaport, ``{Position Location for Futuristic Cellular
  Communications: 5G and Beyond},'' \emph{IEEE Communications Magazine},
  vol.~59, no.~1, pp. 70--75, 2021.

\bibitem{Keating2019}
R.~Keating, M.~Saily, J.~Hulkkonen, and J.~Karjalainen, ``{Overview of
  positioning in 5G new radio},'' \emph{Proceedings of the International
  Symposium on Wireless Communication Systems}, vol. 2019-Augus, pp. 320--324,
  2019.

\bibitem{9473676}
X.~Gao, Y.~Liu, and X.~Mu, ``{SLARM: Simultaneous Localization and Radio
  Mapping for Communication-aware Connected Robot},'' in \emph{2021 IEEE
  International Conference on Communications Workshops (ICC Workshops)}, 2021,
  pp. 1--6.

\bibitem{4103919}
D.~J. Torrieri, ``{Statistical Theory of Passive Location Systems},''
  \emph{IEEE Transactions on Aerospace and Electronic Systems}, vol. AES-20,
  no.~2, pp. 183--198, 1984.

\bibitem{7880669}
M.~Koivisto, M.~Costa, J.~Werner, K.~Heiska, J.~Talvitie, K.~Lepp{\"{a}}nen,
  V.~Koivunen, and M.~Valkama, ``{Joint Device Positioning and Clock
  Synchronization in 5G Ultra-Dense Networks},'' \emph{IEEE Transactions on
  Wireless Communications}, vol.~16, no.~5, pp. 2866--2881, 2017.

\bibitem{Koivisto2017}
M.~Koivisto, A.~Hakkarainen, M.~Costa, J.~Talvitie, K.~Heiska, K.~Leppanen, and
  M.~Valkama, ``{Continuous high-accuracy radio positioning of cars in
  ultra-dense 5G networks},'' \emph{2017 13th International Wireless
  Communications and Mobile Computing Conference, IWCMC 2017}, pp. 115--120,
  2017.

\bibitem{Rastorgueva-Foi2019}
E.~Rastorgueva-Foi, M.~Costa, M.~Koivisto, J.~Talvitie, K.~Leppaneny, and
  M.~Valkama, ``{Beam-based Device Positioning in mmWave 5G Systems under
  Orientation Uncertainties},'' \emph{Conference Record - Asilomar Conference
  on Signals, Systems and Computers}, vol. 2018-Octob, pp. 3--7, 2019.

\bibitem{Zhang2018}
X.~Zhang, S.~M. Razavi, F.~Gunnarsson, K.~Larsson, J.~Manssour, M.~Na, C.~Choi,
  and S.~Jo, ``{Beam-based vehicular position estimation in 5G radio access},''
  \emph{IEEE Wireless Communications and Networking Conference, WCNC}, vol.
  2018-April, pp. 1--6, 2018.

\bibitem{Abu-Shaban2020}
Z.~Abu-Shaban, H.~Wymeersch, T.~D. Abhayapala, and G.~Seco-Granados,
  ``{Single-anchor two-way localization bounds for 5G mmWave systems},''
  \emph{IEEE Transactions on Vehicular Technology}, vol.~69, no.~6, pp.
  6388--6400, 2020.

\bibitem{8761910}
J.~Talvitie, M.~Koivisto, T.~Levanen, M.~Valkama, G.~Destino, and H.~Wymeersch,
  ``{High-Accuracy Joint Position and Orientation Estimation in Sparse 5G
  mmWave Channel},'' in \emph{ICC 2019 - 2019 IEEE International Conference on
  Communications (ICC)}, 2019, pp. 1--7.

\bibitem{Mendrzik2019a}
R.~Mendrzik, H.~Wymeersch, G.~Bauch, and Z.~Abu-Shaban, ``{Harnessing NLOS
  Components for Position and Orientation Estimation in 5G Millimeter Wave
  MIMO},'' \emph{IEEE Transactions on Wireless Communications}, vol.~18, no.~1,
  pp. 93--107, 2019.

\bibitem{Wymeersch2018}
H.~Wymeersch, N.~Garcia, H.~Kim, G.~Seco-Granados, S.~Kim, F.~Went, and
  M.~Fr{\"{o}}hle, ``{5G mm Wave Downlink Vehicular Positioning},'' \emph{2018
  IEEE Global Communications Conference, GLOBECOM 2018 - Proceedings}, no.~1,
  2018.

\bibitem{nazari2022mmwave}
M.~A. Nazari, G.~Seco-Granados, P.~Johannisson, and H.~Wymeersch, ``{MmWave 6D
  Radio Localization with a Snapshot Observation from a Single BS},''
  \emph{arXiv preprint arXiv:2204.05189}, 2022.

\end{thebibliography}
